\documentclass[onecollarge,natbib]{svjour2}
\bibpunct{[}{]}{;}{n}{}{,} 
\smartqed  
\usepackage{graphicx}
\usepackage{amsmath}				
\usepackage{amssymb}				

\newcommand{\vecperp}[2] {\mathbf{#1}_{\perp}^{#2}}

\newcommand{\Dmeson}{D^{-}}
\newcommand{\pimeson}{\pi^{-}}
\newcommand{\mylambda}{\Lambda_{c}^{+}}
\newcommand{\myprime}[2]{#1^{\prime #2}}
\newcommand{\process}{\pimeson ~p \rightarrow \Dmeson ~\mylambda}
\newcommand{\mybar}[2]{\bar{#1}^{#2}}
\def\bra#1{\mathinner{\langle{#1}|}}
\def\ket#1{\mathinner{|{#1}\rangle}}

\def\vd{{\bf \Delta}_\perp}

\journalname{Few Body Systems}
\begin{document}

\title{A perturbative QCD approach to $\process$
}


\author{Stefan Kofler \and
        Wolfgang Schweiger
}


\institute{S. Kofler \and W. Schweiger \at
              Institut f\"ur Physik, FB Theoretische Physik, Universit\"at Graz, Austria\\
               \email{stefan.kofler@edu.uni-graz.at}\\
              \email{wolfgang.schweiger@uni-graz.at}           
}

\date{Received: date / Accepted: date}

\maketitle

\begin{abstract}
We employ the generalized parton picture to analyze the reaction $\process$. Thereby it is assumed that the process amplitude factorizes into one for the perturbatively calculable subprocess $\bar{u}\ u\rightarrow \bar{c}\ c$ and hadronic matrix elements that can be parameterized in terms of generalized parton distributions for the $\pi^-\rightarrow D^-$ and $p\rightarrow \Lambda_c^+$ transitions, respectively. Representing these parton distributions in terms of valence-quark light-cone wave functions for $\pi$, $D$, $p$ and $\Lambda_c$ allows us to make numerical predictions for unpolarized differential and integrated cross sections as well as spin observables. In the kinematical region where this approach is supposed to work, i.e. $s\gtrsim 20$~GeV$^2$ and in the forward hemisphere, the resulting cross sections are of the order of nb. This is a finding that could be of interest in view of plans to measure $\process$, e.g., at J-PARC or COMPASS.

\end{abstract}

\section{Motivation}\label{sec:motivation}
Exclusive production of charmed hadrons is still a very controversial topic. Experimental data are very scarce and theoretical predictions differ by orders of magnitude, depending on the approach used. From general scaling considerations~\cite{Brodsky:1987} one expects, e.g., that the $\bar{p}\,p\rightarrow \bar{\Lambda}_c^-~\Lambda_c^+$ cross section is suppressed by at least two to three orders of magnitude as compared to the $\bar{p}~p \rightarrow \bar{\Lambda}~\Lambda$ cross section. This means that one probably has to deal with cross sections of the order of nb, a challenge which nevertheless seems to be experimentally treatable, as the measurement of $e^+ e^- \rightarrow \Lambda_c^+ ~\bar{\Lambda}_c^-$ cross sections has shown~\cite{Pakhlova:2008vn}. A considerable improvement of the experimental situation on pair production of charmed hadrons is to be expected from the $\bar{\mathrm P}$ANDA detector at FAIR~\cite{Lutz:2009ff}. Another class of reactions for which experimental data may become available even in the near future is the pion-induced exclusive production of charmed hadrons as planned, e.g., at J-PARC~\cite{Shirotori:2015eqa}.

\begin{figure}
\centering
\includegraphics[width=0.5\textwidth]{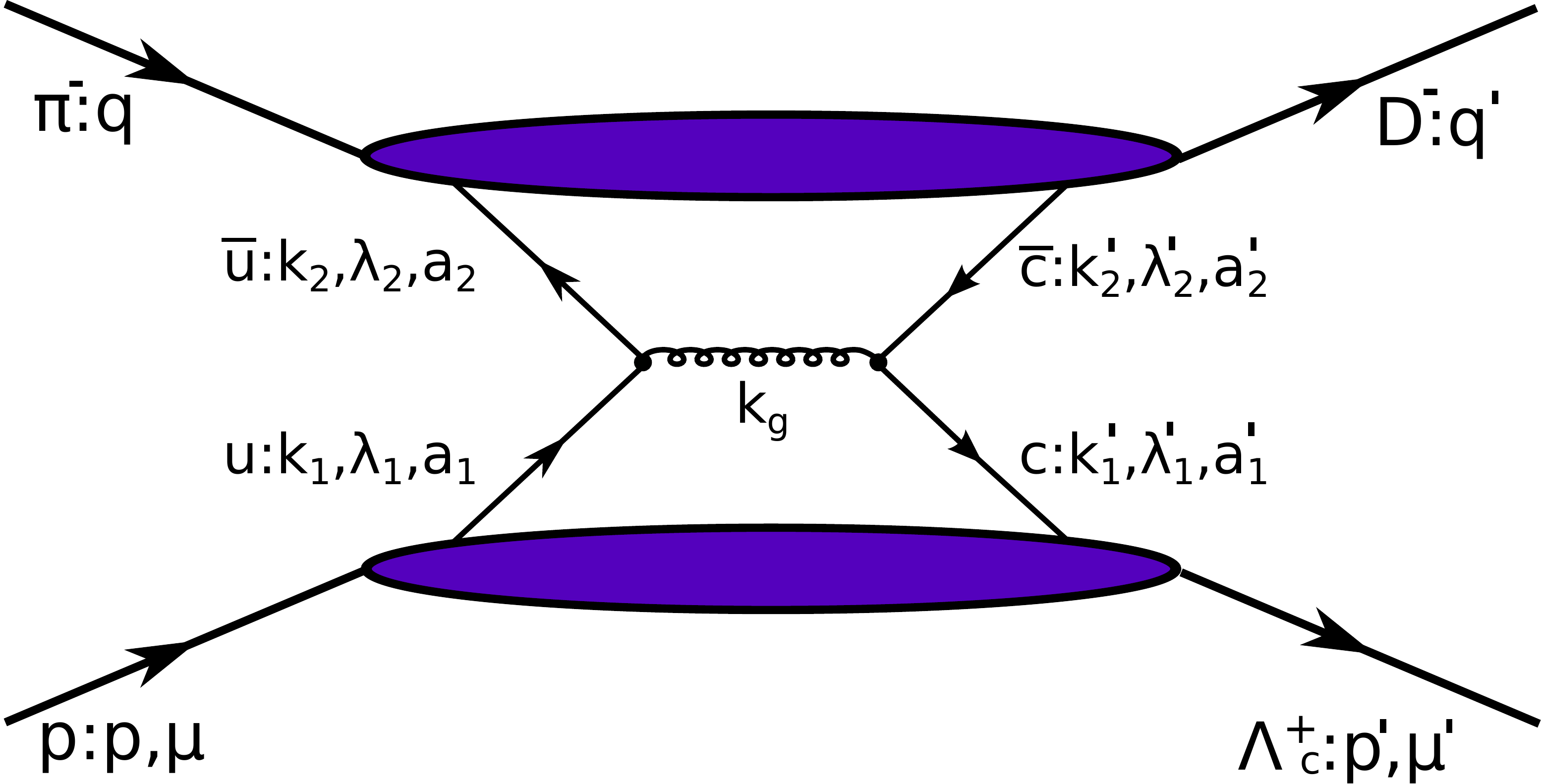}
\caption{The double-handbag contribution to $\process$ (with momenta, LC-helicities and color of the individual particles indicated).}
\label{fig:kinematics}       
\end{figure}

In the present contribution we are going to present a theoretical analysis of $\process$ based on the generalized parton picture. Assuming the intrinsic charm of the $p$ (and the $\pi$) to be negligible, the charmed hadrons in the final state are produced via a handbag-type mechanism (see Fig.~\ref{fig:kinematics}). The blobs in Fig.~\ref{fig:kinematics} indicate soft hadronic matrix elements that are parameterized in terms of generalized parton distributions (GPDs), the $\bar{c}$-$c$ pair is produced perturbatively with the $c$-quark mass $m_c$ acting as a hard scale. A model for the $p\rightarrow \Lambda_c^+$ GPDs is available from foregoing work on $\bar{p}~p\rightarrow \bar{\Lambda}_c^-~\Lambda_c^+$~\cite{Goritschnig:2009sq}. The new ingredients are the $\pi^- \rightarrow D^-$ transition GPDs which we model in analogy to Ref.~\cite{Goritschnig:2009sq} as overlap of valence-quark light-cone wave functions for $\pi^-$ and $D^-$. With these models for the $p\rightarrow \Lambda_c^+$ and  $\pi^- \rightarrow D^-$ GPDs we will estimate the contribution of our handbag-type mechanism to the $\process$ cross section. In Sec.~\ref{sec:formalism} we will sketch the steps and assumptions which finally give us a factorized form of the hadronic $\process$ scattering amplitude. Here also the model wave functions leading to the GPDs used for the numerical calculations are presented. Section~\ref{sec:results} contains the numerical predictions and a short discussion of competing production mechanisms based on hadrondynamics and our conclusions.

\section{Double-handbag amplitude, GPDs and transition form factors}\label{sec:formalism}
We consider $\process$ in a symmetric CM frame. This means that the transverse component of the momentum transfer $\Delta=(p^\prime-p)=(q-q^\prime)$ is symmetrically shared between the particles and the 3-vector part of the average momentum $\bar{p}=(p+p^\prime)/2$ is aligned along the $z$-axis (for assignments of momenta see Fig.~\ref{fig:kinematics}). Expressed in light-cone coordinates the particle four-momenta can then be written as
\begin{align}
\begin{aligned}
\label{eq_mom1}
p &= \Bigg[ (1 + \xi) \mybar{p}{+}, \frac{m_{p}^{2} + \vecperp{\Delta}{2}/4}{2(1 + \xi) \bar{p}^{+}},-\frac{\vecperp{\Delta}{}}{2} \Bigg], \quad\, q  = \Bigg[ \frac{m_{\pi}^{2} + \vecperp{\Delta}{2}/4}{2(1 + \eta) \bar{q}^{-}},~ (1 + \eta) \mybar{q}{-}, ~~\frac{\vecperp{\Delta}{}}{2} \Bigg], \\
\myprime{p}{} &= \Bigg[ (1 - \xi) \mybar{p}{+}, \frac{M_{\Lambda_c}^{2} +  \vecperp{\Delta}{2}/4}{2(1 - \xi) \mybar{p}{+}},\frac{\vecperp{\Delta}{}}{2} \hspace{0.07cm} \Bigg], \quad
\myprime{q}{}  =  \Bigg[ \frac{M_{D}^{2} + \vecperp{\Delta}{2}/4}{2(1-\eta)\mybar{q}{-}},(1-\eta)\mybar{q}{-}, -\frac{ \vecperp{\Delta}{}}{2} \Bigg],
\end{aligned}
\end{align}
where we have introduced the skewness parameter $\xi=-\Delta^+/2\bar{p}^+$. The minus component of the average momentum $\bar{q}^-=(q^-+q^{\prime-})/2$ and the skewness parameter $\eta=\Delta^-/2\bar{q}^-$ have been introduced for convenience, but are determined by $\xi$, $\bar{p}^+$ and $\vecperp{\Delta}{2}$.

The hadronic amplitude as depicted in Fig.~\ref{fig:kinematics} can then be written in the form
\begin{eqnarray}
{\cal M} &=&
\int d^4 k_1^{\mathrm{av}} \theta(k_1^{\mathrm{av}+})
                    \int \frac{d^4 z_1}{(2\pi)^4} e^{ik_1^{\mathrm{av}} z_1}
                    \int d^4 k_2^{\mathrm{av}} \theta(k_2^{\mathrm{av}-})
                    \int \frac{d^4 z_2}{(2\pi)^4} e^{ik_2^{\mathrm{av}} z_2} \nonumber\\[0.2em]
    &\times& \langle \Lambda_c^+\!\!: p^\prime\mu^\prime| T\, \bar{\Psi}^c(-z_1/2)
             \Psi^u(z_1/2)|p\! :p\mu\rangle\;
             \widetilde{H}
             (k_1^{\prime}, k_2^{\prime};k_1,k_2)\;
             \langle D^-\!\!: q^\prime\nu^\prime| T\, \bar{\Psi}^u(z_2/2)
                    \Psi^c(-z_2/2)|\pi^-\!\!:q\nu\rangle\, , \nonumber\\
       \label{eq:ampl}
\end{eqnarray}
with $\tilde {H}$ denoting the perturbatively calculable kernel that represents the partonic subprocess $\bar{u}~u \rightarrow g \rightarrow \bar{c}~c$.
Here two of the four integrations over the quark 4-momenta (and the corresponding Fourier transforms) have already been eliminated by introducing average quark momenta $k_i^{\mathrm{av}}=(k_i+k_i^\prime)/2$ and the fact that the momentum transfer on hadron and parton level should be the same, i.e. $(k_1-k_1^\prime) =(p-p^\prime)$ and $k_2-k_2^\prime=(q-q^\prime)$ (a consequence of translational invariance). The further analysis of $\mathcal{M}$ makes use of the collinear approximation for the active partons. This means that their momenta are replaced by
\begin{align}
\begin{aligned}
{k}_1 &\approx \left[\;{k}_1^+\,,\, \frac{x_1^2
\vd^2}{8{k}_1^+}\,,
                     -\frac12 x_1\vd\right]\,, \quad {k}^\prime_1 \approx\left[\;{k}_1^{\prime +},
                      \frac{m_c^2+x_1^{\prime 2}\vd^2/4}
                      {2{k}_1^{\prime +}}, \frac12 x_1^\prime\vd \right],
\\[0.1em]
{k}_2 &\approx \left[\;\frac{x_2^2\vd^2}{8{k}_2^-}\,,
                    \,{k}_2^-\,, \phantom{-}\frac12 x_2\vd\right],\quad
{k}^\prime_2
\approx \left[\frac{m_c^2+x_2^{\prime 2}\vd^2/4}{2{k}_2^{\prime -}},
                                           {k}_2^{\prime -}, -\frac12 x_2^\prime\vd\right]\, ,
\end{aligned}
\end{align}
with the momentum fractions $x_i^{(\prime)}$ defined by $k_1^{(\prime) +}=x_1^{(\prime)} p^{(\prime) +}$ and $k_2^{(\prime) -}=x_2^{(\prime)} q^{(\prime) -}$. For later purposes it is also convenient to introduce average momentum fractions $\bar{x}_i$ via the relations $x_1^{(\prime)}=(\bar{x}_1\pm\xi)/(1\pm\xi)$ and $x_2^{(\prime)}=(\bar{x}_2\pm\eta)/(1\pm\eta)$. The justification of the collinear approximation rests on the physically plausible assumptions that the parton virtualities and (intrinsic) transverse momenta are restricted by a typical hadronic scale of the order of $1$~GeV and that the GPDs exhibit a pronounced peak at large values of $\bar{x}_1$ ($\bar{x}_2$) close to the ratio $\bar{x}_{10}=m_c/M_{\Lambda_c}$ ($\bar{x}_{20}=m_c/M_D$) (for details, see Ref.~\cite{Goritschnig:2009sq}).
As a consequence of the collinear approximation $\tilde{H}$ does not depend on $k_1^{\mathrm{av}-}$, $\vec{k}_{1\perp}^{\mathrm{av}}$, $k_2^{\mathrm{av}+}$, $\vec{k}_{2\perp}^{\mathrm{av}}$ and the corresponding integrations can be carried out leading to delta functions in the associated $z$-variables. These delta functions force the products  of the quark-field operators onto the light cone ($z_1\rightarrow z_1^-$ and $z_2\rightarrow z_2^+$) and the time ordering can be dropped~\cite{Diehl:1998sm}.

To proceed further one picks out the \lq\lq leading twist\rq\rq\ contributions from the bilocal quark-field operator products:
\begin{eqnarray}
\bra{\mylambda}\overline{\Psi}^{c}(-z_1^-/2)  \Psi^{u}(z_1^-/2) \ket{p} &:& \bra{\mylambda} \overline{\Psi}^{c}(-z_1^-/2) \big \{\gamma^{+}\, , \gamma^{+}\gamma_5 , i\sigma^{+j} \big \} \Psi^{u}(z_1^-/2) \ket{p}\, , \label{eq_operator_twist1}\\
 \bra{\Dmeson}\overline{\Psi}^u(z_2^+/2) \Psi^c(-z_2^+/2)\ket{\pimeson} &:& \bra{\Dmeson}  \overline{\Psi}^u(z_2^+/2) \big \{ \gamma^{-}, \gamma^{-} \gamma_5, i\sigma^{-j} \big \} \Psi^c(-z_2^+/2) \ket{\pimeson}\, . \label{eq_operator_twist2}
\end{eqnarray}
The three Dirac structures showing up in Eqs.~(\ref{eq_operator_twist1}) and (\ref{eq_operator_twist2}) can be considered as $+$ or $-$ components of (bilocal) vector, pseudovector and tensor currents, respectively.
These currents are then Fourier transformed (with respect to $z_1^-$ or $z_2^+$) and decomposed into appropriate hadronic covariants. The  coefficients in front of these covariants  are the quantities which are usually understood as GPDs.
Due to parity invariance the matrix elements \( \bra{\Dmeson} \overline{\Psi}^u \gamma^{-} \gamma_5 \Psi^c \ket{\pimeson} \) vanish and the covariant decomposition of the remaining vector and tensor currents gives rise to two \(\pimeson \rightarrow \Dmeson \) transition GPDs,  \( H^{\overline{cu}}_{\pi D} \) and \( E_{T \pi D}^{\overline{cu}} \), which are defined by:
\begin{equation}
\begin{split}
\label{eq_mesonic_GPDs}
\bar{q}^{-} \int \frac{d z^{ +}_{2}}{2\pi} e^{i \bar{x}_{2} \bar{q}^{-}z^{+}_{2}}
                \bra{\Dmeson:q^\prime }   \overline{\Psi}^{u}(z^{+}_{2} /2) \Big \{ \gamma^- , i \sigma^{-j} \Big \} \Psi^{c}(-z^{+}_{2}/2) \ket{\pimeson:q} \\
              \hspace{2.0cm} = \Big \{ 2 \mybar{q}{-}\, H^{\overline{cu}}_{\pi D} (\bar{x}_{2},\eta,t), \frac{\mybar{q}{-} \Delta^{j} -  \Delta^{-} \mybar{q}{j} }{m_\pi + M_D} E_{T \pi D}^{\overline{cu}} (\bar{x}_{2},\eta,t)  \Big \}.
\end{split}
\end{equation}
An analogous analysis leads to eight GPDs for the $p\rightarrow \Lambda_c^+$ transition~\cite{Goritschnig:2009sq}. These are functions of $\bar{x}_1$, $\xi$ and $t=\Delta^2$.

Having expressed the soft hadronic matrix elements in terms of generalized parton distributions one ends up with an integral in which these parton distributions, multiplied with the hard partonic scattering amplitude $H_{\lambda_1^\prime \lambda_2^\prime ,\lambda_1 \lambda_2}\left(\bar{x}_{1} \bar{p}^{+},\bar{x}_{2} \bar{q}^{-}\right)$, are integrated over $\bar{x}_1$ and $\bar{x}_2$. The fact that a heavy $\bar{c}\, c$ pair has to be produced (since we neglect non-perturbative intrinsic charm in the light hadrons) means that the virtuality of the intermediate gluon should be larger than $4 m_c^2\approx 6.3$~GeV$^2$. This justifies the perturbative treatment of $\bar{u}\, u\rightarrow \bar{c}\, c$ and puts (for fixed $s > (M_{\Lambda_c}+M_D)^2\approx 17.27$~GeV$^2$) kinematical constraints on $\bar{x}_1$ and $\bar{x}_2$. For $s$ well above the production threshold ($s \gtrsim 20$~GeV$^2$) and in the forward-scattering hemisphere it can be checked numerically that these constraints imply $\bar{x}_1 > \xi$ and $\bar{x}_2 > \eta$. This means that only the DGLAP region of the GPDs  ($|\bar{x}_1| >  \xi$, $|\bar{x}_2| > \eta$) can contribute to our handbag-type mechanism, an important observation which simplifies the modeling of the GPDs.

The supposition that the  \( p \rightarrow \Lambda_c^+ \) and \( \pimeson \rightarrow \Dmeson \) GPDs are strongly peaked at $\bar{x}_{10}$ and $\bar{x}_{20}$, respectively, leads to a further simplification of the $\process$ amplitude.
The major contributions to the $\bar{x}_1$ and $\bar{x}_2$ integrals will then come from $\bar{x}_1\approx \bar{x}_{10}$ and $\bar{x}_2\approx \bar{x}_{20}$.
One can thus replace the hard partonic scattering amplitude by its value at the peak position, $H_{\lambda_1^\prime \lambda_2^\prime ,\lambda_1 \lambda_2}\left(\bar{x}_{10} \bar{p}^{+},\bar{x}_{20}\bar{q}^{-}\right)$ and take it out of the integral.
What one is left with are separate integrals over the GPDs which may be interpreted as generalized \( p \rightarrow \Lambda_c^+ \)  and \(\pimeson \rightarrow \Dmeson \) transition form factors.
With this \lq\lq peaking approximation\rq\rq\ our final expressions for the $\process$ amplitudes become:
\begin{align}
\label{eq_procamp}
 &\mathcal{M}_{+,+} =\mathcal{M}_{-,-} =\frac{1}{4}  \sqrt{1 - \xi^2} \, H_{+-,+-} \, R_V \, G\, , \nonumber \\
&\mathcal{M}_{+,-} =  -\mathcal{M}_{-,+}  =\frac{1}{4}   \sqrt{1 - \xi^2} \,  H_{++,-+}\, S_T \, G,
\end{align}
\noindent with the \(\pimeson \rightarrow\Dmeson \) transition form factor
\begin{equation}\label{eq_piDff}
G(\eta,t) = \int_{\eta}^{1} \frac{\mathrm{d} \bar{x}_2 }{\sqrt{\bar{x}^{2}_2 - \eta^2}}
H^{\overline{cu}}_{\pi D}(\bar{x}_2,\eta,t) \, .
\end{equation}
In Eqs.~(\ref{eq_procamp}) we have restricted ourselves to the two most important $p\to\Lambda_c$ GPDs, $H^{cu}_{p\Lambda_c}$ and $H^{cu}_{T p\Lambda_c}$ (these are associated with $\gamma^+$ and $\sigma^{+j}$). The respective form factors are $R_V$ and $S_T$, defined analogously to Eq.~(\ref{eq_piDff}).
The underlying assumption is that those GPDs (and corresponding form factors) which involve non-zero orbital angular momentum of the (anti)quarks that make up the hadrons are suppressed. This leads also to omission of \( E_{T \pi D}^{\overline{cu}}\).
\begin{figure}[t!]
\centering
  \includegraphics[width=0.45\textwidth]{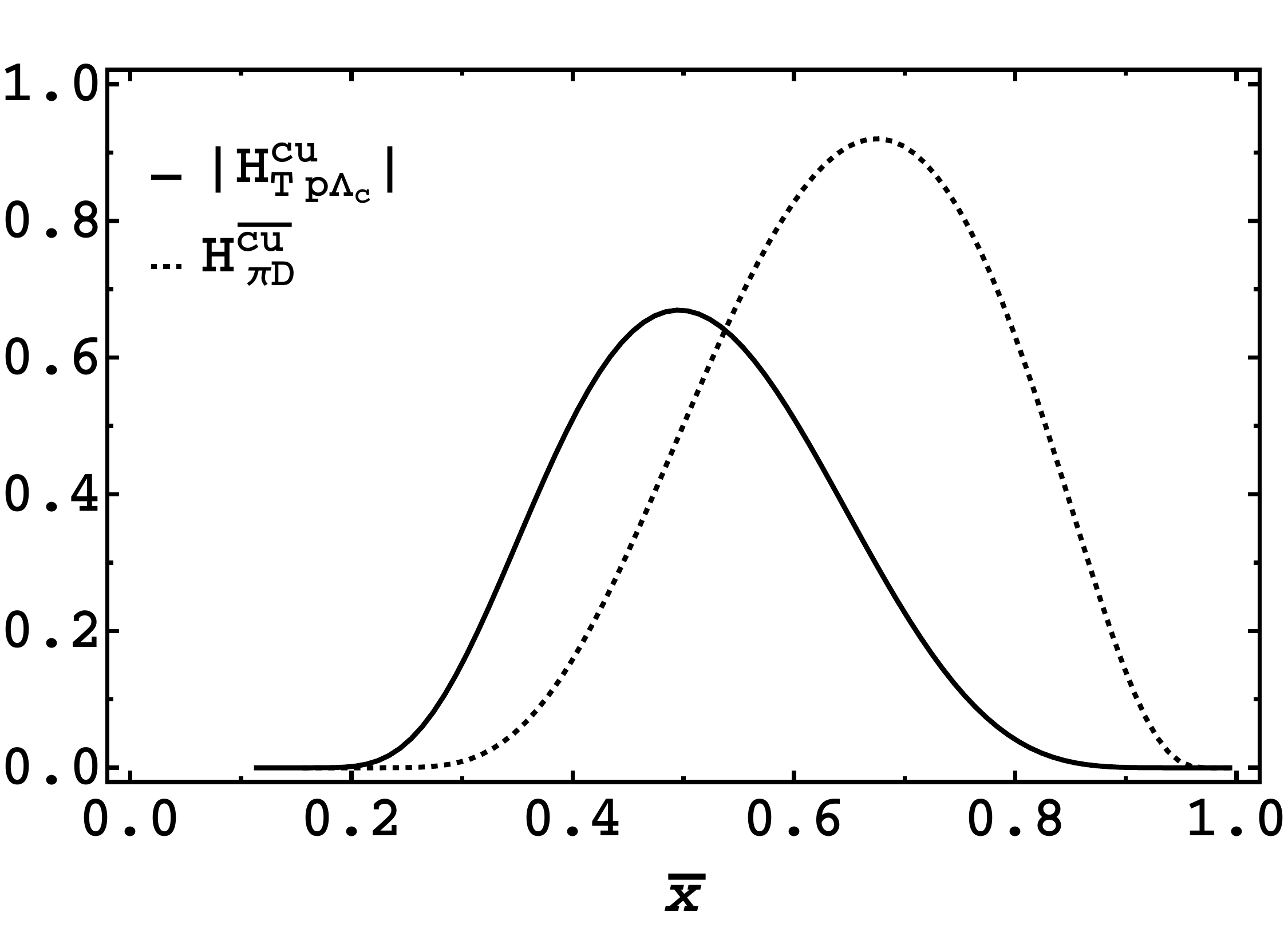}\hspace{0.05\textwidth}
  \includegraphics[width=0.45\textwidth]{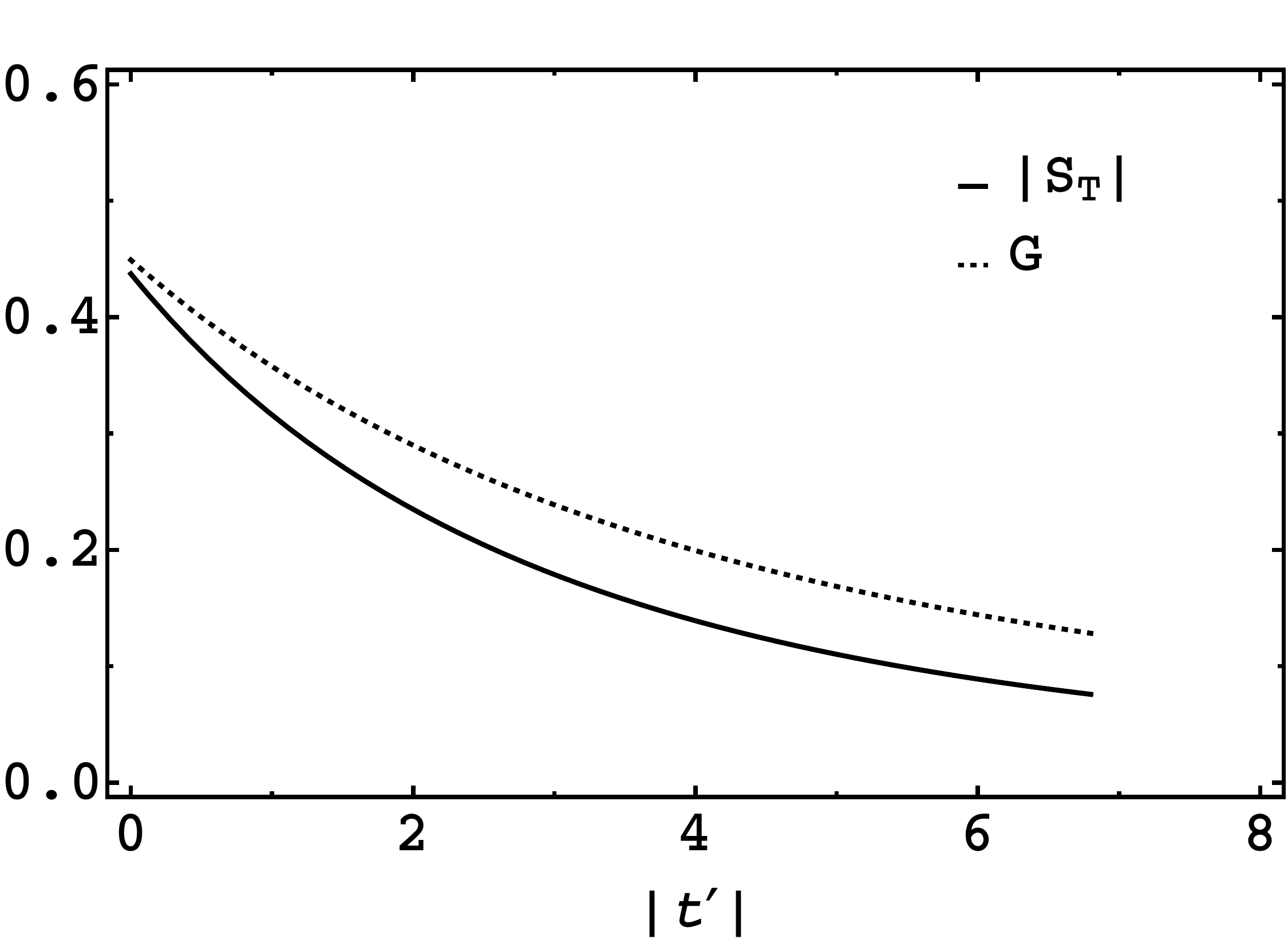}
\caption{Left panel: The $p\rightarrow \Lambda_c^+$ (solid line) and $\pi^-\rightarrow D^-$ (dotted line) transition GPDs $H^{cu}_{T\, p\Lambda_c}$ and $H^{\overline{cu}}_{\pi D}$ obtained with the wave functions (\ref{eq:ppiwfs}) and (\ref{eq:DLwfs}) and the $KK$ mass exponential for $s=25$~GeV$^2$ and $\vec{\Delta}_\perp^2=0$.
 Right panel: The corresponding transition form factors $|S_T|$ and $G$ (see Eq.~(\ref{eq_piDff})) as functions of $|t^\prime|=|t-t_0|$, where $t_0$ is the $t$-value for forward scattering.}
\label{fig:gpds}       
\end{figure}

This concludes the general analysis of our process. The next step is the modeling of the GPDs. As mentioned already above, we have to consider only the DGLAP region There it is possible to model the GPDs as overlaps of light-cone wave functions for the valence Fock states of the respective hadrons~\cite{Diehl:2000xz}. For the pion and the proton we take the parameterisations of the light-cone wave functions proposed in Ref.~\cite{Bolz:1996sw} and Ref.~\cite{Feldmann:1999sm}
\begin{equation}
\label{eq:ppiwfs}
\psi_{\pi} \left ( x, {\mathbf{k}}_{\perp} \right)= N_\pi \, \exp \left[ \frac{-a_{\pi}^{2} \, {\mathbf{k}}_{\perp}^{ 2} }{{x} (1-{x})}\right]\, , \qquad
\psi_{p}({x}_i,{\mathbf{k}}_{\perp\, i})= N_p (1+3x_1)
         \exp{\left[-a_p^2\sum{\frac{{\mathbf{k}}_{\perp\, i}^2}{{x}_i}}\right]}\,.
\end{equation}
This forms are supported by several phenomenological applications. Similar wave functions are taken for the $D$ and the $\Lambda_c$:
\begin{equation}
\label{eq:DLwfs}
\psi_{D} ( {x}, {\mathbf{k}}_{\perp})= N_D \, \exp\left[ -f(x) \right] \, \exp \left[ \frac{-a_{D}^{2} \, {\mathbf{k}}_{\perp}^{2} }{{x} (1-{x})}\right]\, , \,
\psi_{\Lambda_c}({x}_i,{\mathbf{k}}_{\perp\, i})= N_{\Lambda_c} \exp\left[ -f(x_1) \right] \,
         \exp{\left[-a_{\Lambda_c}^2\sum{\frac{{\mathbf{k}}_{\perp\, i}^2}{{x}_i}}\right]}\,.\\
\end{equation}
The mass exponential generates the expected peak at $x_{(1)} \approx \bar{x}_{i0}$ (with $x_{(1)}$ being the momentum fraction of the heavy quark). The parameters $N_D$ and $a_D$ are chosen such that the experimental value of the D-meson decay constant $f_D=0.207$~GeV is reproduced and the valence-Fock-state probability becomes $0.9$~\cite{Kofler:2014yka}. An appropriate choice of parameters for the $\Lambda_c$ wave function can be found in Ref.~\cite{Goritschnig:2009sq}, where an overlap representation of $p\rightarrow \Lambda_c$ GPDs has been derived and applied to $\bar{p}\,p\rightarrow \bar{\Lambda}_c^-~\Lambda_c^+$. We use two types of mass exponentials that have been suggested in the literature \cite{Korner:1992uw,Ball:2008fw}:
\begin{equation}
f_{KK} (x) = \frac{a_{\Lambda_c (D))}^2 M_{\Lambda_c (D)}^2 \left(x - \bar{x}_{i0} \right)^2}{x (1-x)}\, ,\quad
f_{BB} (x) =a_{\Lambda_c (D)}^2 M_{\Lambda_c (D)} (1-x)\, .
\end{equation}

For a more detailed account of the formalism and the modeling of the GPDs we refer to Refs.~\cite{Goritschnig:2009sq,Kofler:2014yka}, where also analytical formulae for the model GPDs, hard scattering amplitudes, etc. can be found. An impression how our model GPDs and corresponding form factors for the $\pi^-\rightarrow D^-$ and $p\rightarrow\Lambda_c^+$ transitions look like can be gained from Fig.\ref{fig:gpds}. The general behavior of the GPDs is that with increasing $s$ the maximum moves to larger values of $\bar{x}$ and becomes smaller. It should also be mentioned that the two $p\rightarrow\Lambda_c^+$ form factors (and corresponding GPDs) that show up in Eq.~(\ref{eq_procamp}) are approximately the same for reasonably small probability ($\lesssim 10\%$) to find the $c$-quark with helicity opposite to the $\Lambda_c$ helicity in the $\Lambda_c$~\cite{Goritschnig:2009sq}.

\section{Results and discussion}\label{sec:results}
\begin{figure}[t!]
\centering
  \includegraphics[width=0.45\textwidth]{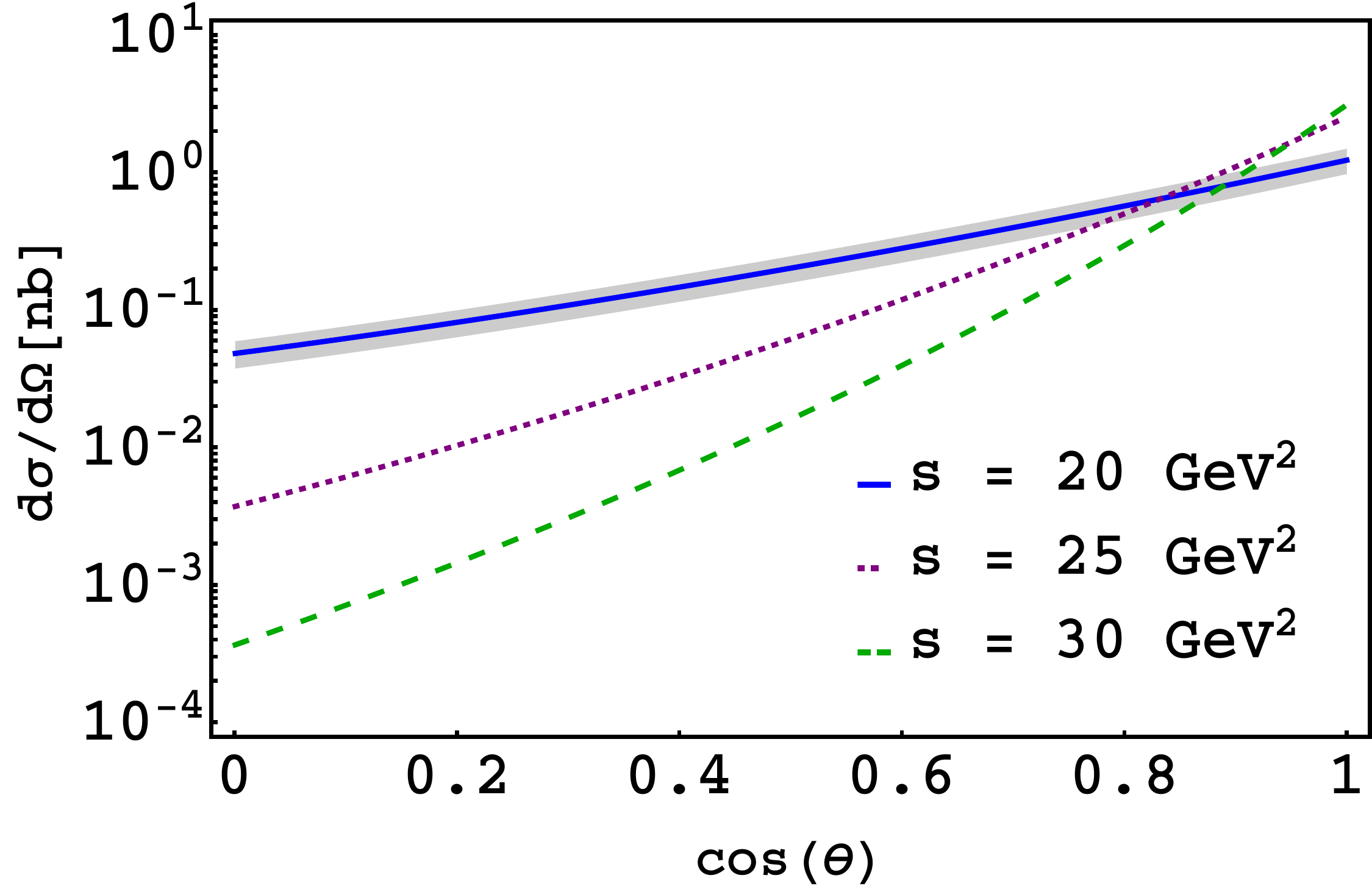}\hspace{0.05\textwidth}
  \includegraphics[width=0.45\textwidth]{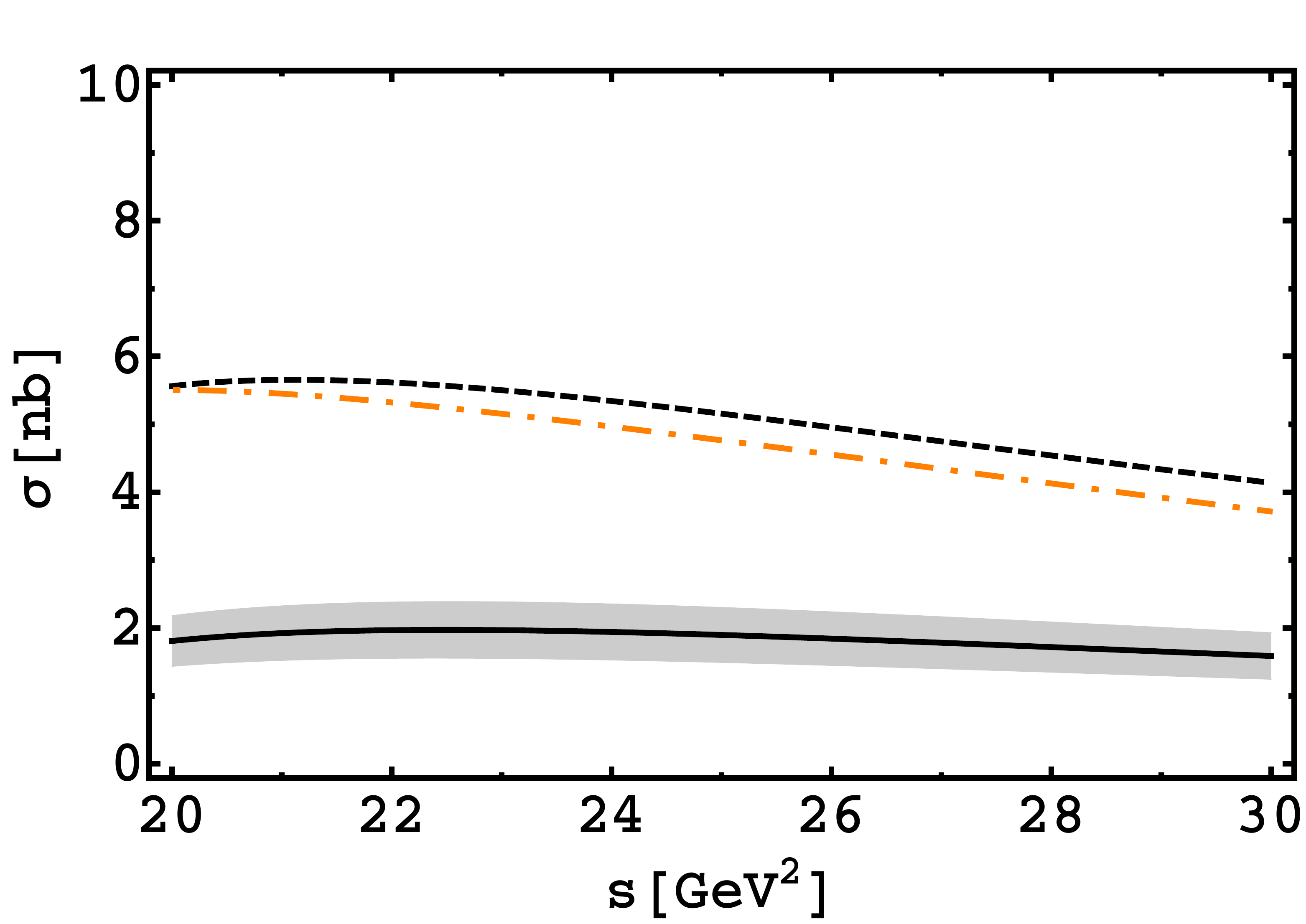}
\caption{Left panel: The differential $\process$ cross section in the CM system versus $\cos\theta$ obtained with the KK mass exponential for $s=20, 25, 30$~GeV$^2$. Right panel: The integrated $\process$ cross section versus Mandelstam $s$ for the $KK$ (solid line) and the $BB$ (dashed line) mass exponential in comparison with predictions from a hybrid Regge model~\cite{Kim:2016imp} (dash-dotted line). The effects of uncertainties in the $\Lambda_c$ and $D^-$ wave-function parameters (in case of the KK mass exponential) are indicated by the shaded band.}
\label{fig:cross-secs}       
\end{figure}

With these model GPDs (and corresponding form factors) we are now able to calculate the hadronic scattering amplitudes (see Eq.~(\ref{eq_procamp})) and cross sections. Predictions for unpolarized differential and integrated cross sections obtained with the KK mass exponential are displayed in Fig.~\ref{fig:cross-secs}. The shaded band exhibits the variation of the cross section, if the wave-function parameters are varied within a reasonable range. Integrated cross-section results are also shown for the BB mass exponential. Obviously the differences between predictions obtained with different analytical forms of the wave functions are larger than the variations coming from parametric errors in the wave functions. The integrated cross sections are of the order of nb with the BB mass exponential giving the larger results.
This is the order of magnitude that has also been found for $ \bar{p}\, p \rightarrow  \bar{\Lambda}_c^-\, \Lambda_c^+ $~\cite{Goritschnig:2009sq} and $ \bar{p}\, p \rightarrow \bar{D}^0 D^0$~\cite{Goritschnig:2012vs}, when treated within the generalized parton framework. It seems to be in accordance with old AGS experiments at  \( s \approx 25~\mbox{GeV}^2 \) which found upper bounds of \( 7~\mbox{nb} \) for \( \pi^-\, p\rightarrow D^{\ast -} \, \Lambda_c^+ \) and \( \approx 15~\mbox{nb} \) for \( \process \)~\cite{Christenson:1985ms}.

Interestingly, an integrated cross section comparable to ours has been found by the authors of Ref.~\cite{Kim:2016imp} (dash-dotted line in Fig.~\ref{fig:cross-secs}), who used a hybridized Regge model, i.e. a production mechanism mainly based on hadron dynamics. This is in contrast to the findings for $ \bar{p}\, p \rightarrow  \bar{\Lambda}_c^-\, \Lambda_c^+ $ and $ \bar{p}\, p \rightarrow \bar{D}^0 D^0$, where models using single~\cite{Haidenbauer:2016pva,Haidenbauer:2014rva} or Reggeized hadron exchange~\cite{Titov:2008yf,Khodjamirian:2011sp} provide cross sections which are two to three orders of magnitude larger than those from the handbag-type mechanism. The crucial point in these models seems to be the strength of the $D^{(\ast)}pY_c$ coupling which is either fixed by $SU(4)$-flavor symmetry~\cite{Haidenbauer:2016pva,Haidenbauer:2014rva,Titov:2008yf} or by QCD sum rules~\cite{Khodjamirian:2011sp}. In our case this coupling corresponds to the form factors $G$ and $S_T$. These are the quantities which make the cross section  that small. The flavor dependence of the wave functions leads to a strong mismatch of light and heavy hadron wave functions in the overlap integral. If this mismatch would not exist, our cross section would be about two orders of magnitude larger, as we have seen in a similar calculation of $ \bar{p}\, p \rightarrow  \bar{\Lambda}_c^-\, \Lambda_c^+ $ performed within a quark-diquark model~\cite{Kroll:1988cd}.

Cross sections as large as predicted by the hadronic models would also indicate that, in contrast  to our assumption, charm is produced non-perturbatively which means that (non-perturbative) intrinsic charm of the proton must be taken into account. This could, in principle, be done within our approach. Then the charmed hadrons in the final state would, in addition to the handbag mechanism, be produced via mechanisms which are fed by the ERBL region of the GPDs. It is, however, hardly conceivable that the small amount of intrinsic charm in the proton that is compatible with inclusive data~\cite{Jimenez-Delgado:2014zga} could increase the cross section for the exclusive production of charmed hadrons by two or three orders of magnitude. This holds in particular for the kinematic situations we are considering, where the skewness parameter and thus also the ERBL region becomes small.
Experimental data for processes like $ \process $,  $\bar{p} ~p \rightarrow \bar{\Lambda}^-_c ~\mylambda$, $\gamma\, p\rightarrow \bar{D}^0\, \Lambda_c^+$ and $\bar{p} ~p \rightarrow \bar D^0~{D}^0$ up to several GeV above production threshold would thus be highly desirable to pin down the production mechanism of charmed hadrons and shed some more light on the question of non-perturbative intrisic charm in the proton.

\begin{acknowledgements}
S. Kofler acknowledges the support of the Fonds zur F\"orderung der wissenschaftlichen Forschung in \"Osterreich (Grant No. FWF DK W1203-N16). We are thankful to S.-H. Kim for providing us with the numerical data of his results that are shown in Fig.~\ref{fig:cross-secs}.
\end{acknowledgements}



\end{document}